# Fast $O_{expected}(N)$ Algorithm for Finding Exact Maximum Distance in $E^2$ Instead of $O(N^2)$ or $O(N \lg N)$

Vaclav Skala

*Department of Computer Science and Engineering, Faculty of Applied Sciences, University of West Bohemia, CZ 306 14 Plzen, Czech Republic*

**Abstract.** This paper describes novel and fast, simple and robust algorithm with $O(N)$ expected complexity which enables to decrease run-time needed to find an exact maximum distance of two points in $E^2$. The proposed algorithm has been evaluated experimentally on larger different datasets. The proposed algorithm gives a significant speed-up to applications, when medium and large data sets are processed. It is over 10 000 times faster than the standard algorithm for $10^6$ points randomly distributed points in $E^2$. Experiments proved the advantages of the proposed algorithm over the standard algorithm and convex hull diameters approaches.

## INTRODUCTION

A maximum distance of two points in the given data set is needed in many applications. A standard "Brute Force" algorithm with $O(N^2)$ complexity is usually used, where $N$ is a number of points in the given data set. Such algorithm leads to very high run-time if larger data sets are to be processed. Typical data sets in computer graphics contain usually $10^5$-$10^7$ and even more of points. In spite of the CPU speed increases, the run-time even for such a simple task leads to unacceptable processing time.

However our task is just to find the maximum distance, not all the pairs having a maximum distance. Therefore the complexity of this algorithm should be lower. Also due to the numerical precision points do not lie exactly on a circle if data have this very specific property.

## BRUTE FORCE ALGORITHM

The standard "Brute Force" (BF) algorithm uses two nested loops in order to find a maximum distance. Algorithms with such approach can be found in many textbooks dealing with fundamental algorithms and data structures, e.g. [4], [5], [7], [8], [12]. Such algorithms can be represented by Alg.1 in general as:

```
function distance_2 (A , B: point);
{  distance_2:=(A.x-B.x)^2 + (A.y-B.y)^2};
# Square of the distance ‖A-B‖ #

d := 0;
for i := 1 to N-1 do
 for j := i+1 to N do
  {
    d0 := distance_2(X_i , X_j);
    if d < d0 then d := d0
  };
d := SQRT (d)  # if needed #
```

**ALGORITHM 1**. "Brute Force" algorithm

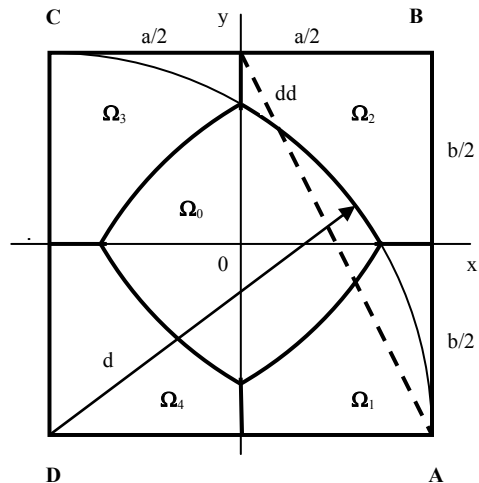

**FIGURE 1.** Splitting the $\Omega_0$ set to $\Omega_i$ sets of the worst case

The Alg.1 is clearly of $O(N^2)$ complexity and processing time increases significantly with number of points processed, see Tab.1.

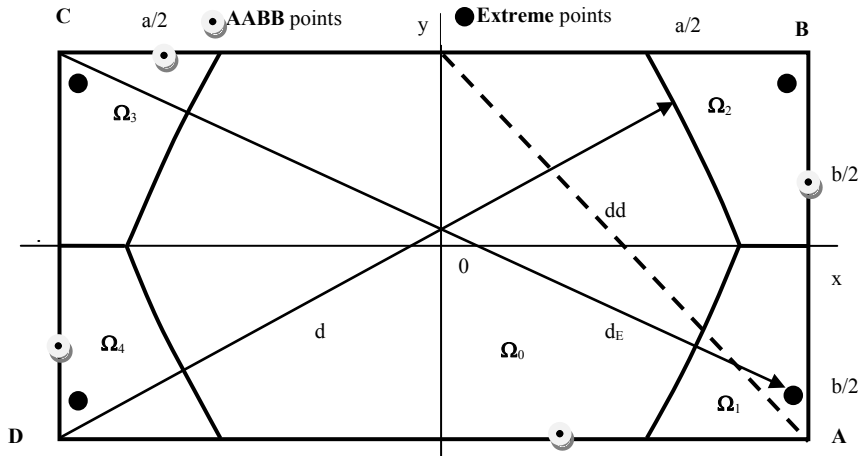

**FIGURE 2.** Splitting the $\Omega_0$ set to $\Omega_i$ sets for the rectangular area case

In practice, it can be expected that points are not organized in a very specific manner, e.g. points on a circle etc., and points are uniformly distributed more or less. In this case "output sensitive" algorithms usually lead to efficient solutions.

Let points are inside of an Axis Aligned Bounding Box (AABB) defined as $<-a/2, a/2> \times <-b/2, b/2>$. Fig.1 presents a typical situation for the worst case when AABB is a square ($a = b$), while the Fig.2 presents general AABB situation for the case $a > b$. In the following we will explore the worst case, i.e. situation at the Fig.1, and the first maximum distance estimation $d$ is $d = a$.

It can be seen that points in the set $\Omega_0$ cannot influence the maximum distance computation in the given data set. We can remove all points $\Omega_0$ from the given data set $\Omega$ and obtain faster algorithm. As the maximum distance finding algorithm is of $O(N^2)$ complexity, an algorithm with a lower complexity can be used in order to find and eliminate points which cannot influence the final distance. Space subdivision techniques can be used to split points into the disjunctive data sets $\Omega_i$ and decrease run-time complexity again. For a general case, when AABB is not squared, the $\Omega_0$ set will contain more points of course, see Fig.2.

Let us explore the worst case more in a detail and consider the case, when $a = b$.

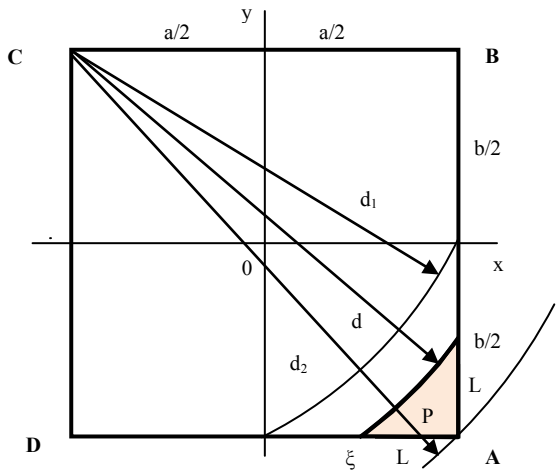
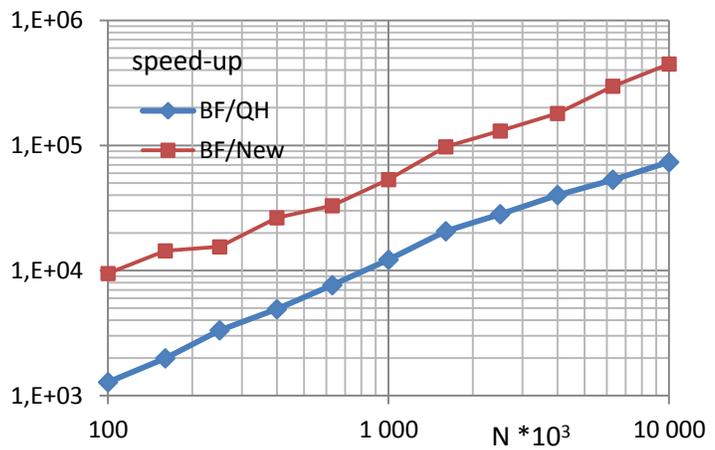

**FIGURE 3**. Distances definition for a general AABB

**FIGURE 4.** Experimental speed up of the proposed algorithm; convex hull comparison is included

Radiuses $d_1$ and $d_2$ are given as $d_1^2 = [\max\{a,b\}]^2 + \frac{1}{2}[\min\{a,b\}]^2$, $d_2^2 = a^2 + b^2$. An extension to $E^3$ is straightforward and easy to implement.

# PROPOSED ALGORITHM

The new proposed algorithm was developed for larger data sets and it is based on "in-core" technique, i.e. all data are stored in a computer memory. The fundamental requirements for the algorithm development were: simplicity, robustness and simple extensibility to $E^3$. The proposed algorithm is based on two main principles:
- Remove as many non-relevant points as possible
- Divide and conqueror technique in order to decrease algorithm complexity

Fig.1 shows five regions $\Omega_i$, where the given points are located. It should be noted that the worst case is presented, i.e. when AABB is a square. Let us assume that the points that cannot contribute to the final maximum distance are located in the region $\Omega_0$, which contains points closer to all corners of the AABB than the *minimal* edge length of the AABB or known distance estimation. Then the given data set $\Omega$ can be reduced to $\Omega = \Omega - \Omega_0$. In order to decrease expected number of points to be processed, we need to process this data set $\Omega$ to get more information on those points.

The following principal steps have to be made:
1. Pre-processing: can be performed with *O(N)* complexity:
   a. Find the bounding AABB and extreme points, i.e. two extreme points for each axis (max. 4 points).
   b. Find the most distant "extreme" points [max] for each corner of the AABB (max. 4 points).
   c. Find the minimum distant "extreme" points [min] for each corner of the AABB (max. 4 points).
   d. Determine the longest mutual distance *d* between those found points (max.12 points). It should be noted that the worst case is a squared AABB and found distance $d \geq a$. For a rectangular window found distance $d \geq max\{a,b\}$.
   e. Determine points of $\Omega_0$ that cannot contribute to the maximum distance, i.e. points having a smaller distance than the found distance *d* from all corners of the AABB and extreme points. Remove the $\Omega_0$ points from the original data set $\Omega$.
   f. Split remaining points to new sets $\Omega_i$, i=1,..,4, see the Fig.1.
2. Run-time steps of the proposed algorithm:
   a. Taking an advantage of space subdivision, find the maximum distance *d* between points of [$\Omega_1$, $\Omega_3$], i.e. one point from $\Omega_1$ and the second point is from $\Omega_3$ as there can be expected the longest distance between the given points - this step is $O(N^2)$ complexity.
   b. Remove points from the $\Omega_2$ and $\Omega_4$ datasets closer to the related corner of the AABB than already found distance *d* - this step is *O(N)* complexity.
   c. Find a new maximum distance *d* between points of [$\Omega_2$, $\Omega_4$] - this step is $O(N^2)$ complexity.
   d. If already found distance $d \leq dd$ then
      i. Reduce $\Omega_1$, $\Omega_2$, $\Omega_3$, $\Omega_4$ - steps are $O(N^2)$ complexity
      ii. find a new maximum distance *d* between points of [$\Omega_1$, $\Omega_2$], [$\Omega_2$, $\Omega_3$], [$\Omega_3$, $\Omega_4$] and [$\Omega_4$, $\Omega_1$]. It is necessary to note that if $d > dd$, then the $\Omega_0$ boundary crosses the AABB and points in the neighbors regions cannot contribute to the maximum distance.

As can be seen the algorithm is very simple and easy to implement.

**Implementation notes**

There are several possibilities how to further improve the proposed algorithm especially in the context of the specific programming language and data structures used. However, the influence of this is small as experiments proved and for the expected data sizes do not have any significant influence. Generally, it is recommended:
- The "array list" construction should be used for storing $\Omega_i$ sets; this construction enables to increase an array size without reallocation and data copying,
- Two or higher dimensional arrays for storing x, y values should not be used, as for each array element one addition and one multiplication operations are needed (computational cost is hidden in the index evaluation). Data should be stored in two arrays **X** and **Y**, or as pairs (x, y) in one-dimensional structure array **XY** etc.
- Square of a distance should be used in order to save multiple square root evaluations. It is possible as the square function is monotonically growing and can be used for comparison operations.

It should be noted that the $\Omega_i$ sets are determined by an arc of a circle, i.e. the separation function is quadratic. Experiments proved that if a half-space separation function is used as in Fig.2, the proposed algorithm is faster as only a linear function is evaluated.

## EXPERIMENTAL RESULTS

Standard PC with 2,8 GHz Intel Pentium 4, 1 GB RAM with MS Windows XP was used. Cumulative results obtained by experiments are presented in Tab.1. Experiments proved that the speed-up grows significantly with the number of points processed. Delphi/Pascal was used for the experimental implementation.

**TABLE 1**. Experimental results for uniformly distributed points [* values obtained by extrapolation]

| Points | Computational time | | Speed-up [* times] |
|---|---|---|---|
| $10^3*N$ | Brute Force (BF) [s] | New algorithm [s] | BF/New |
| 1 000 | 13 784 | 0.26 | 53 277 |
| 4 000 | 221 674 | 1.23 | 180 094 |
| 10 000 | 1 106 173* | 2.47 | 447 916 |

Detailed experimental comparison of the proposed algorithm with the standard and convex hull diameter algorithms is presented by Fig.4. It can be seen that the speed-up of the proposed algorithm is significant, e.g. for $10^7$ points speed up is approx. **4,5 $10^7$ times** and grows with number of points processed. There might be additional speed up if GUP is used for processing.

## CONCLUSION

A new simple, easy to implement, robust and effective algorithm for finding a maximum distance of points in $E^2$ was developed and experimentally verified. The experimental results clearly proved that the proposed algorithm is convenient for medium and large data sets. Algorithm speed-up grows significantly with the number of points processed. The proposed algorithm can be easily extended to $E^3$ by a simple modification. In the $E^3$ case, we have to process $\Omega_1, \ldots, \Omega_8$ data subsets. However the memory requirements and preprocessing time remain the same as we have to only split data from the $\Omega$ set to the $\Omega_i$ datasets which are smaller, but number of processed points remains constant.

## ACKNOWLEDGMENTS

The author would like to express his thanks to students of the University of West Bohemia, especially to students V.Ondracka, V.Bystricky, P.Kellnhofer, V.Gersl, O.Petrik, Z.Prokop, B.Podlesak, J.Vyskovsky for experimental verification of the proposed algorithm and an extensive testing. Many thanks are given to anonymous reviewers for their valuable comments and suggestions that significantly contributed to the improvement of this paper. This work was supported by the Ministry of Education of the Czech Republic – projects LG13047, LH12181.